\pdfoutput=1
\documentclass[11pt]{article}
\usepackage{amsmath}
\usepackage{graphicx,epstopdf}
\usepackage{cite}
\usepackage{array}
\usepackage{graphicx}
\usepackage[font={small}]{caption}
\usepackage{xcolor}
\usepackage[colorlinks=true,
            linkcolor=red,
            urlcolor=green,
            citecolor=blue]{hyperref}          

\catcode`@=12
\definecolor{RawSienna}{cmyk}{0,0.72,1,0.45}
\definecolor{dgreen}{rgb}{0.0,0.42,0.13}
\definecolor{darkblue}{rgb}{0.0, 0.0, 0.55}
\definecolor{cornellred}{rgb}{0.7, 0.11, 0.11}
\definecolor{calpolypomonagreen}{rgb}{0.08, 0.5, 0.5}
\newcommand{\R}{\color{RawSienna}}
\newcommand{\G}{\color{dgreen}}
\newcommand{\B}{\color{darkblue}}

\def\beq{\begin{equation}}
\def\eeq{\end{equation}}
\def\bea{\begin{eqnarray}}
\def\eea{\end{eqnarray}}
\makeatletter
\@addtoreset{equation}{section}

\begin{document}
\title{\LARGE \bf Evaluation of the Majorana Phases of a General Majorana Neutrino Mass Matrix: Testability of hierarchical Flavour Models }
\author{{\bf Rome Samanta\footnote{ rome.samanta@saha.ac.in}, Mainak Chakraborty\footnote{mainak.chakraborty@saha.ac.in}, Ambar Ghosal\footnote{ambar.ghosal@saha.ac.in},}\\
Saha Institute of Nuclear Physics, 1/AF Bidhannagar,
  Kolkata 700064, India\\ } 
\maketitle
\begin{abstract}
We evaluate the Majorana phases for a general $3\times3$ complex symmetric neutrino mass matrix on the basis of Mohapatra-Rodejohann's phase convention using the three rephasing invariant quantities $I_{12}$, $I_{13}$ and $I_{23}$ proposed by Sarkar and Singh. We find them interesting  as they allow us to evaluate each Majorana phase in a model independent way even if one eigenvalue is zero. Utilizing the solution of a general complex  symmetric mass matrix for eigenvalues and mixing angles we determine the Majorana phases for both the hierarchies, normal and inverted, taking into account the constraints from neutrino oscillation global fit data as well as bound on the sum of the three light neutrino masses ($\Sigma_im_i$) and the neutrinoless double beta decay ($\beta\beta_{0\nu}$) parameter $|m_{11}|$. This  methodology of finding the Majorana phases is applied thereafter in some predictive models for both the hierarchical cases (normal and inverted) to evaluate the corresponding Majorana phases and it is shown that all the sub cases presented in inverted hierarchy section can be realized in a model with texture zeros and scaling ansatz within the framework of inverse seesaw although one of the sub case following the normal hierarchy is yet to be established. Except the case of quasi degenerate neutrinos, the methodology obtained in this work is able to evaluate the corresponding Majorana phases, given any model of neutrino masses.
\end{abstract}
\newpage
\section{Introduction}
Apart from hierarchical structure of massive neutrinos a fundamental qualitative nature of these elusive particles whether they are Dirac or Majorana type is yet unknown. Neutrinoless double beta decay ($\beta \beta_{0\nu}$) mode \cite{{Serra:2014saa,Agostini:2013mzu,Majorovits:2015vka,::2015uaa,Agostini:2015fga,Xu:2015dfa,Simkovic:2012hq,Gomez-Cadenas:2015twa,Tornow:2014vta,Bilenky:2014uka,Hennecke:1975zz,Zuber:2015sxa} 
 } is able to discriminate between the two different types. Positive evidence of the above experimental search will be able to determine the Majorana nature of neutrinos assuming the above decay is mediated due to light neutrino. Several $\beta \beta_{0\nu}$ experiments are ongoing and planned. In Ref. \cite{Schwingenheuer:2012zs} a brief discussion about some of the important experiments is presented. Among them, EXO-200\cite{Auger:2012ar} experiment puts an upper limit on the relevant neutrino mass matrix element $|m_{11}|$ within a range as $|m_{11}|<$ (0.14-0.35 eV). Further, NEXT-100 \cite{DavidLorcafortheNEXT:2014fga} experiment will be able to bring down the above value of the order of 0.1 eV. Thus in an optimistic point of view such property of neutrino could be testified by the next generation experiments. However, even if it is possible to pin down the value of $|m_{11}|$, it is still difficult to predict the values of the Majorana phases until we can fix the absolute neutrino mass scale. It is shown in Ref.\cite{Xing:2013woa} that in addition to the $\beta \beta_{0\nu}$ decay experiments, lepton number violating processes in which the Majorana phases show up are also corroborative to determine the individual Majorana phases. Another interesting physical aspect such as contribution of the Majorana phases to the generation of $\theta_{13}$ within the present $3\sigma$ range of neutrino oscillation global fit data is also studied in the literature\cite{Gupta:2014lwa}. Ref. \cite{Minakata:2014jba} discusses how to constrain the Majorana phases using the results from cosmology and double beta decay. Thus it is worthwhile to study the calculability of the Majorana phases in terms of a general neutrino mass matrix ($m_\nu$) parameters.
In the present work we evaluate individual Majorana phases in terms of the  parameters of a general $m_\nu$ using three rephasing invariants $I_{12}$, $I_{13}$ and $I_{23}$ presented in Ref.\cite{Sarkar:2006uf} on the basis of Mohapatra-Rodejohann's phase convention\cite{sc2}. Although there are several papers which discusses the general procedure for calculating the Majorana phases, motivation behind taking the rephasing invariants is that the methodology we present here is capable of calculating the Majorana phase in a model independent way even if one of the eigenvalue is zero which is still allowed as far as the present  neutrino oscillation global fit data is concerned. Moreover as one of the rephasing invariant ($I_{23}$) is directly proportional to $m_3$, therefore it vanishes if $m_3=0$ and hence shows a strong dependency of the Majorana phases with the light neutrino masses. In the present work we evaluate the Majorana phases for a general complex symmetric neutrino mass matrix ($m_\nu$) taking into account the global fit oscillation data and the upper bound on the sum of the three light neutrino masses ($\Sigma_im_i$) along with the $\beta\beta_{0\nu}$ decay parameter for both the hierarchical cases. We then conclude except the case of quasi degeneracy\footnote{For the quasi-degenerate case the procedures of calculating the Majorana phases are stated in Sec. \ref{qsi}}, the methodology presented in this work is able to calculate the Majorana phases, given any model of neutrino masses and for convenience,  we further numerically estimate the ranges of each Majorana phase for both types of hierarchies,  in the context of a cyclic symmetric model as well as a model with scaling ansatz property. It is also shown that all the sub cases we present in inverted hierarchy section of the general discussion can be realized through the choice of a model with scaling ansatz with texture zeros within the framework of inverse seesaw while one of the phenomenologically viable sub case of the normal hierarchy section is yet to be identified. The plan of the paper is as follows.
\subparagraph{}
In Section \ref{s2} we briefly discuss the basic formalism to set the convention of the Majorana phase representation within the framework of neutrino oscillation phenomena. CP violating rephasing invariants are presented in Section \ref{s3}. Section \ref{s4} contains explicit calculation of the Majorana phases  for both types of neutrino mass hierarchies along with phenomenologically viable different sub cases. Numerical estimation of the Majorana phases, their connection to the physical observables and discussions about their testability for the general case taking into account the constraints from the extant data for both types of neutrino mass hierarchies are presented in Section \ref{s5}. In Section \ref{s6} application of the above methodology in the context of  cyclic symmetric and scaling ansatz invariant models is presented. Section \ref{s7} contains summary of the present work.    
\section{Basic formalism}\label{s2}
Experimental observation of neutrino flavour oscillation constitutes a robust evidence in favour of nonzero neutrino masses. The flavour transition process is basically a quantum mechanical interference phenomena with the explicit relationship between the left handed quantum fields ($\nu_{\alpha L}$) of the flavour basis and the mass basis ($\nu_{i L}$) as
\bea
\nu_{\alpha L}&=&\Sigma_i U^*_{\nu \alpha i}\nu_{iL}
\eea
where $\alpha(=1,2,....,m)$ corresponds to the flavour index and $i(=1,2,....,n)$ implies the mass index and the matrix $U_{\nu}$ is the corresponding neutrino mixing matrix. For three generation of fermions, i.e, for $n=m=3$, the weak Lagrangian containing charged lepton fields and the neutrino fields can be written in the mass basis as
\bea
-\mathcal{L}^{cc}&= \frac{g}{\sqrt{2}}&\bar{l}_{\alpha L}\gamma^\mu (U_l^\dagger U^*_\nu)_{\alpha i} \nu_{iL}W_{\mu}^{-}+ h.c.
\eea
where $U_{l}$ is the unitary mixing matrix in the charged lepton sector. The matrix $U_l^\dagger U_\nu$ is the leptonic mixing matrix and is known as the $Pontecorvo-Maki-Nakagawa-Sakata$ mixing matrix ($U_{PMNS}$) which contains 3 mixing angles and 6 phases in general. It is useful to redefine the mixing matrix by absorbing the unphysical phases into the charged lepton fields and the neutrino fields (Dirac type). If the neutrinos are Majorana type, they break the global $U(1)$ symmetry and hence, redefinition of the neutrino fields is not possible. Therefore, out of 6 phases 3 unphysical phases can be absorbed by redefining only the charged lepton fields and thus the $U_{PMNS}$ matrix is parametrized as
\bea
U_{PMNS}&=&U_{CKM}P_M\label{a}
\eea
where $U_{CKM}$ is the usual CKM type matrix and is given by
\bea
U_{CKM}=
\begin{pmatrix}
c_{1 2}c_{1 3} & s_{1 2}c_{1 3} & s_{1 3}e^{-i\delta}\\
-s_{1 2}c_{2 3}-c_{1 2}s_{2 3}s_{1 3} e^{i\delta }& c_{1 2}c_{2 3}-s_{1 2}s_{1 3} s_{2 3} e^{i\delta} & c_{1 3}s_{2 3}\\
s_{1 2}s_{2 3}-c_{1 2}s_{1 3}c_{2 3}e^{i\delta} & -c_{1 2}s_{2 3}-s_{1 2}s_{1 3}c_{2 3}e^{i\delta} & c_{1 3}c_{2 3}
\end{pmatrix}
\eea
where $c_{ij}\Rightarrow \cos \theta_{ij}$,$s_{ij}\Rightarrow \sin \theta_{ij}$ and $\delta$ is the Dirac CP phase. $P_M$ is a $3\times3$ diagonal phase matrix and following Mohapatra-Rodejohann's convention \cite{sc2} it is given by
\bea
P_M&=&(1,e^{i\alpha}, e^{i(\beta+\delta)})
\eea
where $\alpha$ and $\beta +\delta $ are the Majorana phases which do not appear in the $neutrino\rightarrow neutrino$  oscillation experiments \cite{Bilenky:1980cx,Giunti:2010ec}. Regarding the structure of $P_M$ matrix we would like to mention the following: The advantage of using the above Majorana phase convention is that for $m_3=0$ it is  possible to calculate the single existing Majorana phase $\alpha$ while, for $m_1=0$, only the phase difference $\alpha-(\beta+\delta)$ is calculable. The result will be reversed if we utilize the PDG\cite{Beringer:1900zz} convention. Explicitly, with PDG convention, if $m_3=0$, only the phase difference is calculable, however if $m_1$ is vanishing it is possible to calculate the existing Majorana phase. Based on PDG convention two of the authors presented a detailed calculation \cite{Adhikary:2013bma} for both the Majorana phases in context of a general $m_\nu$, however, if one of the eigenvalue is zero  which is still allowed by the present neutrino experimental data, it is not possible to calculate individual phases in that case. The above mentioned problem is successfully resolved in the present work.\\

CP violating effect of Majorana phases in $neutrino\rightarrow antineutrino$ oscillation\cite{Xing:2013ty,Xing:2014eia,Delepine:2009qg} and some lepton number violating (LNV) processes are studied in detail in Ref.\cite{Xing:2013woa}. In this work, using the rephasing invariants constructed out of the neutrino mass matrix elements \cite{Sarkar:2006uf} we determine the Majorana phases for two different hierarchical cases.
\section{CP violating phase invariants}\label{s3}
Considering neutrinos as the Majorana fermions in extended standard model one can parametrize  $U_{PMNS}$ with the CP violating phases as given in Eqn.(\ref{a}) where we redefine the charged lepton fields absorbing the unphysical phases of total mixing matrix $U$. Hence, in principle the mixing matrix $U$ can be defined as 
\bea
U\equiv P_\phi U_{PMNS}\label{uckm}
\eea
where $P_\phi$ is a $3\times 3$ diagonal phase (unphysical) matrix and is given by 
\bea
P_\phi &=&diag (e^{i\phi_1},e^{i\phi_2},e^{i\phi_3}).
\eea
 Now, as the low energy neutrino mass matrix is complex symmetric it can be diagonalized as
\bea
U^\dagger m_\nu U^*&=& d_\nu \label{diag} \eea
where\bea
d_\nu &=&diag(m_1,m_2,m_3).
\eea
Substituting Eqn.(\ref{uckm}) in Eqn.(\ref{diag}) we get
\bea
U_{PMNS}^\dagger P_\phi^\dagger m_\nu P_\phi^* U_{PMNS}^*&=& d_\nu
\eea
 alternately \bea P_\phi^\dagger m_\nu P_\phi^* &=& U_{PMNS} d_\nu U_{PMNS}^T \label{m1}.\eea  
Thus $P_\phi$ rotates the mass matrix $m_\nu$ in phase space. Therefore, the rephasing invariants (remain invariant under phase rotation) of $m_\nu$ contain the informations about the CP violating phases. It has been shown explicitly in Ref.\cite{Sarkar:2006uf} that for three generations of neutrinos there are three independent rephasing invariants and are given by
\bea
I_{12}&=&Im[m_{11}m_{22}m_{12}^*m_{21}^*]\nonumber \\
I_{23}&=&Im[m_{22}m_{33}m_{23}^*m_{32}^*]\nonumber \\
I_{13}&=&Im[m_{11}m_{33}m_{13}^*m_{31}^*]\label{pi2}
\eea
where $m_{\alpha \beta}$ is the element of $m_\nu$ at $\alpha \beta$ position with $\alpha,\beta=1,2,3$. Now since the invariants of Eqn.(\ref{pi2}) are independent of phase rotation of $m_\nu$, therefore to evaluate them in terms of mixing angles, CP violating phases and the eigenvalues we can rewrite Eqn.(\ref{m1}) as
\bea  m_\nu  &=& U_{PMNS} d_\nu U_{PMNS}^T \label{m1m}\eea
where without any loss of generality we assume $\phi_i=0$ which corresponds to the structure of $P_\phi$ as $P_\phi=diag(1,1,1)$. Now writing down Eqn.(\ref{m1m}) explicitly one can find the mass matrix elements as
\bea
 m_{11} & =& c_{12}^2 c_{13}^2m_1 +s_{12}^2 c_{13}^2m_2 e^{2 i\alpha}+m_3s_{13}^2e^{-2i\delta +2i(\beta+\delta)}\label{m2}\\
m_{12}  &= &c_{13}\lbrace -m_1(c_{12}s_{12}c_{23}+c_{12}^2s_{13}s_{23} e^{i\delta})\nonumber \\
&&+m_2  e^{2i\alpha}(c_{12}s_{12}c_{23}-s_{12}^2s_{13}s_{23} e^{i\delta})\rbrace +m_3c_{13}s_{13}s_{23}e^{-i\delta+2i(\beta+\delta)} \\
m_{13}&= &c_{13}\lbrace m_1 (c_{12}s_{12}s_{23}-c_{12}^2s_{13}c_{23} e^{i\delta})\nonumber \\
&& -m_2e^{2 i\alpha}(c_{12}s_{12}s_{23}+s_{12}^2s_{13}c_{23} e^{i\delta})\rbrace +m_3c_{13}s_{13}c_{23}e^{-i\delta+2i(\beta+\delta)} \\ 
m_{22} & =&m_1 (s_{12}c_{23}+c_{12}s_{23}s_{13}e^{i \delta})^2\nonumber \\
&&+m_2 e^{2 i\alpha}(c_{12}c_{23}-s_{12}s_{23}s_{13}e^{i \delta})^2+m_3c_{13}^2s_{23}^2e^{2i(\beta +\delta)}\\
m_{23} &=& m_1\lbrace c_{12}s_{12}s_{13}(c_{23}^2-s_{23}^2)e^{i\delta}
+c_{12}^2c_{23}s_{23}s_{13}^2e^{2i\delta}-s_{12}^2s_{23}c_{23} \rbrace \nonumber \\
 &&-m_2 e^{2i\alpha} \lbrace c_{12}s_{12}s_{13}(c_{23}^2-s_{23}^2)e^{i\delta}
 -s_{12}^2c_{23}s_{23}s_{13}^2e^{2i\delta}+c_{12}^2s_{23}c_{23} \rbrace \nonumber \\ &&+ m_3c_{23}s_{23}c_{13}^2e^{2i(\beta+\delta)}\\
 m_{33} &= &m_1(c_{12}c_{23}s_{13}e^{i\delta}-s_{12}s_{23})^2\nonumber \\
 &&+m_2  e^{2i\alpha}(s_{12}c_{23}s_{13}e^{i\delta}+c_{12}s_{23})^2+m_3c_{23}^2c_{13}^2e^{2i(\beta+\delta)}.\label{m3}
\eea
It is now straightforward to calculate $I_{12}$ and $I_{13}$ using Eqn.(\ref{m2}) to Eqn.(\ref{m3}). Neglecting terms $O(s_{13}^2)$ and higher order we obtain $I_{12}$ and $I_{13}$ as 
\bea
I_{12}&=&A c_{23}^3[Bc_{23}-2s_{23}s_{13}\lbrace c_{12}^2m_1\Phi_1 + s_{12}^2m_2\Phi_2 \rbrace]\nonumber \\&& +m_3^2c_{12}c_{13}^4s_{12}c_{23}[-2c_{12}^2s_{23}^3c_{13}^2m_1s_{13}A_1+2s_{23}^3c_{13}^2s_{12}^2m_2s_{13}A_2]\nonumber \\&&+m_3c_{12}c_{13}^4s_{12}c_{23}[s_{12}s_{23}^2c_{13}^2m_1c_{12}^3c_{23}A_3+s_{12}^3s_{23}^2c_{13}^2m_2c_{12}c_{23}A_4\nonumber \\ &&-2c_{12}^4s_{23}m_1m_2c_{23}^2s_{13}A_5-2s_{23}m_1m_2s_{12}^4c_{23}^2s_{13}A_5+2c_{12}^2s_{23}s_{12}^2s_{13}c_{23}^2
A_6\nonumber \\&&+2c_{12}^4s_{23}^3c_{13}^2m_1^2s_{13}A_7
+4\cos(2\alpha)c_{12}^2s_{23}^2c_{13}^2m_1m_2s_{12}^2s_{13}A_7+2s_{23}^3c_{13}^2m_2^2s_{12}^4s_{13}A_7]\label{I1}
\eea
\bea
I_{13}&=&A s_{23}^3[Bs_{23}+2c_{23}s_{13}\lbrace c_{12}^2m_1\Phi_1 + s_{12}^2m_2\Phi_2 \rbrace]\nonumber \\&& 
-m_3^2c_{12}c_{13}^4s_{12}s_{23}[-2c_{12}^2c_{23}^3c_{13}^2m_1s_{13}A_1+2c_{23}^3c_{13}^2m_2s_{12}^2s_{13}A_2]\nonumber \\&&+m_3c_{12}c_{13}^4s_{12}s_{23}[c_{12}^3c_{23}^2c_{13}^2m_1s_{12}s_{23}A_3+c_{12}c_{23}^2c_{13}^2m_2s_{12}^3s_{23}A_4\nonumber \\ &&+2c_{12}^4c_{23}m_1m_2s_{23}^2s_{13}A_5+2c_{23}m_1m_2s_{12}^4s_{23}^2s_{13}A_5-2c_{12}^2c_{23}s_{12}^2s_{13}s_{23}^2
A_6\nonumber \\&&-2c_{12}^4c_{23}^3c_{13}^2m_1^2s_{13}A_7
-4\cos(2\alpha)c_{12}^2c_{23}^3c_{13}^2m_1m_2s_{12}^2s_{13}A_7-2c_{23}^3c_{13}^2m_2^2s_{12}^4s_{13}A_7]\label{I2}
\eea
where
\bea
A &=&- c_{12}s_{12}m_1m_2c_{13}^4\label{A}\\
B&=&\sin(2\alpha)c_{12}s_{12}(m_2^2-m_1^2)\label{B}\\
\Phi_1 &=& \lbrace \sin(2\alpha-\delta)m_1+\sin [\delta] m_2\rbrace\\
\Phi_2 &=& \lbrace \sin(2\alpha+\delta)m_2-\sin [\delta] m_1\rbrace
\eea
and
\bea
A_1&=&\sin(\delta)m_1+\sin(2\alpha-\delta)m_2 \nonumber\\
A_2&=&\sin(\delta)m_2-\sin(2\alpha+\delta)m_1 \nonumber\\
A_3&=&\sin 2(\beta+\delta)m_1^2+2\sin(2\alpha-2\beta-2\delta)m_1m_2-\sin(4\alpha-2\beta-2\delta)m_2^2\nonumber \\
A_4&=&\sin 2(\alpha+\beta+\delta)m_1^2-2\sin 2(\beta+\delta)m_1m_2-\sin(2\alpha-2\beta-2\delta)m_2^2\nonumber \\
A_5&=&\sin(2\alpha-2\beta-\delta)m_1+\sin(2\beta+\delta)m_2\nonumber \\
A_6&=&\sin(2\beta+\delta)m_1^3-\sin(2\alpha+2\beta+\delta)m_1^2m_2
-\sin(4\alpha-2\beta-\delta)m_1m_2^2+\sin(2\alpha-2\beta-\delta) \nonumber \\
A_7&=&\sin(2\beta+\delta)m_1+\sin(2\alpha-2\beta-\delta)m_2
.\eea
A careful inspection reveals that the invariants are expressed in a tricky way. To be more precise, they are written as \bea 
I_{ij}&=&\zeta_1 + s_{13}\zeta_2+m_3^2\zeta_3+m_3\zeta_4
\eea
where `$\zeta_i$' is some parameter dictated by Eqn.(\ref{I1}) and Eqn.(\ref{I2}). The reason behind such a way to write down the invariants are the following: firstly, the popular paradigm in the neutrino mass models is to generate vanishing $\theta_{13}$ at the leading order and thereafter nonzero value of the same is generated by the means of some perturbation to the mass matrix and finally as the oscillation data dictates the mass square differences only, there is also a possibility of a vanishing neutrino mass (e.g, models with scaling ansatz, Zee-Babu model etc.). Therefore one can see the direct impact of their presence or absence in the measures of CP violation.\\
Now the remaining invariant ($I_{23}$) has a special character that it vanishes for $m_3=0$ and it comes out as
\bea
I_{23}&=&m_3^3c_{23}s_{23}c_{13}^2B_1+m_3^2c_{23}s_{23}c_{13}^2[2c_{12}^3c_{13}^2m_{2}s_{12}s_{13}(c_{23}^2-s_{23}^2)B_2\nonumber\\
&&-2c_{12}c_{13}^2m_1s_{12}^3s_{13}(c_{23}^2-s_{23}^2)B_3]+m_3c_{23}s_{23}c_{13}^2[c_{12}^6c_{23}m_2^3s_{23}B_4\nonumber \\ &&+c_{12}^4c_{23}m_1m_2^2s_{12}^2s_{23}B_5+c_{12}^2c_{23}^2m_1^2m_2s_{12}^4s_{23}B_6+c_{23}m_1^3s_{12}^6s_{23}B_7\nonumber \\
&&-2c_{12}^5m_2^2s_{12}s_{13}(c_{23}^2-s_{23}^2)B_8-4\cos(2\alpha)c_{12}^3m_1m_2s_{12}^3s_{13}(c_{23}^2-s_{23}^2)B_8]\nonumber \\
&&-2c_{12}m_1^2s_{12}^5s_{13}(c_{23}^2-s_{23}^2)B_8]\label{I3}
\eea
with
\bea
B_1&=&\sin 2(\alpha-\beta-\delta)c_{12}^2c_{23}^2c_{13}^4m_2s_2-\sin 2(\beta +\delta)c_{23}c_{13}^4m_1s_{12}^2s_{23}\nonumber \\
&=&\Phi_1 \nonumber \\
B_2&=&\sin (2 \alpha-\delta)m_1+\sin(\delta)m_2\nonumber \\
&=&-\Phi_2 \nonumber \\
B_3&=& \sin(\delta)m_1-\sin(2 \alpha+\delta)m_2\nonumber \\
B_4&=&-\sin(2 \alpha-2\beta-2\delta)\nonumber \\
B_5&=&2\sin 2(\beta+\delta)-\sin(4\alpha-2\beta-2\delta)\nonumber \\
B_6&=&-2\sin(2 \alpha-\beta-\delta)+\sin 2(2 \alpha+\beta+\delta)\nonumber \\
B_7&=&\sin 2(\beta+\delta)\nonumber \\
B_8&=&\sin(2\beta+\delta)m_1+\sin(2\alpha- 2 \beta- \delta)m_2.
\eea
\section{The Majorana phases}\label{s4}
At the outset, first, we would like to mention that the three independent invariants $I_{12}$, $I_{13}$ and $I_{23}$ stand for the three CP violating phases $\alpha$, $\beta+\delta$ and $\delta$, however in this section we solve the invariants only for the Majorana phases ($\alpha,\beta+\delta$) while the Dirac CP phase $\delta$ is calculable from the usual Jarlskog measure of CP violation. Next, for a general $m_\nu$ where all the parameters are present and all the eigenvalues and mixing angles are nonzero, all the invariants are independent and in principle one can extract the $\alpha$ and $\beta+\delta$ phases without any specific hierarchical assumption which is also useful for the quasi-degenerate case. However, the calculation is too cumbersome in this general situation. In the present work we consider a simplified approach assuming hierarchical structure of neutrino masses and calculate the Majorana phases utilizing the invariants $I_{12}$, $I_{13}$ and $I_{23}$ for both, normal and inverted hierarchical cases. 
\subsection{Inverted hierarchy ($m_2>m_1>>m_3$)}\label{invh}
\textbf{Case I:} $m_1,m_2,m_3\neq 0$, $\theta_{13}\neq 0$: {\B$Three$ $independent$ $invariants$.}\\

\noindent
In this case utilizing Eqn.(\ref{I1}) and (\ref{I2})the Majorana phase $\alpha$ comes out as
\bea
\alpha & =&\frac{1}{2}\sin^{-1}\left\lbrace- \frac{I_{12}s_{23}^2+I_{13}c_{23}^2}{c_{23}^2s_{23}^2c_{13}^4c_{12}^2s_{12}^2m_1m_2\Delta m_\odot ^2}\right\rbrace\label{2.1}\eea
where $\Delta m_\odot ^2=m_2^2-m_1^2$ and  we neglect the terms containing $m_3(m_{min})$ in both the invariants ($I_{12}$ and $I_{13}$). Another equivalent expression of $\alpha$ can also be obtained from Eqn.(\ref{m2})(neglecting the term containing $m_3s_{13}^2$) showing explicit relationship with $\beta\beta_{0\nu}$ decay parameter $|m_{11}|$ as
\bea
\alpha &=& \frac{1}{2}\cos^{-1}\left\lbrace \frac{|m_{11}|^2}{2c_{12}^2s_{12}^2c_{13}^4m_1m_2}-\frac{(c_{12}^4m_1^2+s_{12}^4m_2^2)}{2c_{12}^2s_{12}^2m_1m_2}\right\rbrace. \label{alph}
\eea
In principle we can use any of the equation (Eqn.(\ref{2.1}) or Eqn.(\ref{alph})) to find $\alpha$. The first one depends upon the explicit construction of $I_{12}$ and $I_{13}$ in terms of the neutrino mass matrix ($m_\nu$) elements while the second one requires the knowledge of $\beta\beta_{0\nu}$ decay parameter $|m_{11}|$. \\
In order to calculate $\beta+\delta$ from Eqn.(\ref{I3}) the terms involving $s_{13}(c_{23}^2-s_{23}^2)$ can be neglected. Therefore, assuming inverted hierarchy $I_{23}$ can be approximated with dominant term as
\bea
I_{23}&=&m_2^3m_3c_{23}s_{23}c_{13}^2c_{12}^6c_{23}s_{23}B_4\nonumber \\
&=&-m_2^3m_3c_{23}^2s_{23}^2c_{13}^2c_{12}^6\sin(2\alpha-2[\beta+\delta]).
\eea
Reverting the above equation the Majorana phase $\beta+\delta$ is expressed as
\bea
\beta+\delta &=&-\frac{1}{2}\sin^{-1}\left\lbrace-\frac{I_{23}}{m_2^3m_3c_{23}^2s_{23}^2c_{13}^2c_{12}^6}\right\rbrace +\alpha.\label{c1}
\eea

\noindent
\textbf{Case II:} $m_1,m_2,\theta_{13}\neq 0$, $m_3=0$: {\B$Two$ $independent$ $invariants$.}\\

\noindent
In this case utilizing Eqn.(\ref{I1}),(\ref{I2}) and (\ref{I3}) the three rephasing invariants $I_{12},I_{13}$ and $I_{23}$ come out as
\bea
I_{12}&=&A c_{23}^3[Bc_{23}-2s_{23}s_{13}\lbrace c_{12}^2m_1\Phi_1 + s_{12}^2m_2\Phi_2 \rbrace]\label{m4}\\
&=&I_{12}^0-2Ac_{23}^3s_{23}s_{13}\lbrace c_{12}^2m_1\Phi_1 + s_{12}^2m_2\Phi_2 \rbrace \\
I_{13}&=&A s_{23}^3[Bs_{23}+2c_{23}s_{13}\lbrace c_{12}^2m_1\Phi_1 + s_{12}^2m_2\Phi_2 \rbrace]\label{m5}\\
&=&I_{13}^0+2As_{23}^3c_{23}s_{13}\lbrace c_{12}^2m_1\Phi_1 + s_{12}^2m_2\Phi_2 \rbrace\\
I_{23}&=&0\label{m6}
\eea
where \bea
I_{12}^0&=&ABc_{23}^4\\
I_{13}^0&=&ABs_{23}^4
\eea
with A, B already defined in Eqn.(\ref{A}) and (\ref{B}) respectively. As one of the invariant vanishes due to the condition $m_3=0$, therefore, the three independent CP phases can not be solved from the above invariants and thus the two non zero invariants corresponds to one Majorana phase ($\alpha$) and the Dirac CP phase ($\delta$) as $\beta+\delta$ vanishes for $m_3=0$. Proceeding as previous  we get the same expression for the Majorana phase $\alpha$ as given in Eqn.(\ref{alph}). Furthermore, solving Eqn.(\ref{m4}) to Eqn.(\ref{m5}) an equivalent expression of $\alpha$, same as Eqn.(\ref{2.1}) is also obtained.\\

\noindent
\textbf{Case III:} $m_1,m_2\neq0$, $m_3, \theta_{13}= 0$: {\B$One$ $independent$ $invariant$.}\\

\noindent
In this case  the invariants given in Eqn.(\ref{m4}),(\ref{m5}) and (\ref{m6}) become \bea
I_{12}&=&I_{12}^0\\
I_{13}&=&I_{13}^0
\eea
and \bea I_{23}&=&0 \eea
\noindent
It is amply clear that the first two invariants $I_{12}$ and $I_{13}$ are not independent\cite{sc2} of each other and their correlated relationship leads to the estimation of only one  Majorana phase  $\alpha$ while the information about the Dirac CP phase is lost.
\subsection{Normal hierarchy ($m_3>>m_2>m_1$)}\label{nrmlh}
\textbf{Case I:} $m_1,m_2,m_3\neq 0$, $\theta_{13}\neq 0$: {\B$Three$ $independent$ $invariants$.}\\

\noindent
In this case since $m_1=m_{min}$ and $m_3>>m_2>m_1$, we simplify $I_{12}$ and $I_{13}$ as
\bea
I_{12}&=&\kappa\sin(2\alpha-2[\beta+\delta])+\eta s_{13}s_{23}^2\sin[\delta]\nonumber \\
I_{13}&=&\kappa\sin(2\alpha-2[\beta+\delta])-\eta s_{13}c_{23}^2\sin[\delta]\label{3.1}
\eea
where the parameters $\kappa$ and $\eta$ are defined through \bea \kappa &=&-c_{12}^2c_{23}^2c_{13}^6m_2^3m_3s_{12}^4s_{23}^2\eea
\bea
\eta &=&2c_{12}c_{23}c_{13}^6m_2^2m_3^2s_{12}^3s_{23}.\eea
Now from Eqn.(\ref{3.1}) we get 
\bea
\sin(2\alpha-2[\beta+\delta])&=&\left\lbrace\frac{c_{23}^2 I_{12}+s_{23}^2 I_{13}}{\kappa}\right\rbrace \nonumber \\
&=&\Gamma \label{2.2}
.\eea 
Again due to the hierarchical condition $m_3>>m_2>m_1$, $I_{23}$ can be approximated as 
\bea
I_{23}&\simeq &m_3^3c_{23}s_{23}B_1 \nonumber \\
&=&m_3^3c_{23}s_{23}[\sin (2\alpha-2[\beta+\delta])c_{12}^2c_{23}c_{13}^4m_2s_2-\sin 2(\beta +\delta)c_{23}c_{13}^4m_1s_{12}^2s_{23}].\label{2.3}
\eea
Inserting Eqn.(\ref{2.2}) in (\ref{2.3}) we get 
\bea
\beta+\delta &=&\frac{1}{2}\sin^{-1}\left\lbrace\frac{m_2}{m_1}ct_{12}^2\Gamma-\frac{I_{23}}{m_3^3m_1c_{23}^2s_{23}^2s_{12}^2c_{13}^6}\right\rbrace
\eea
where $ct_{12}\Rightarrow \cot \theta_{12}$.\\
It is now straight forward to calculate the other Majorana phase $\alpha$ from Eqn.(\ref{2.2}) and it comes out as
\bea
\alpha &=&\frac{\sin^{-1}\Gamma + 2(\beta + \delta)}{2}.\label{c2}
\eea
\textbf{Case II:} $m_2,m_3,\theta_{13}\neq 0$, $m_1=0$: {\B$Two$ $independent$ $invariants$.}\\

\noindent
In this case neglecting terms like $s_{13}^2$ and $s_{13}(c_{23}^2-s_{23}^2)$ in $I_{23}$ only the Majorana phase difference $(\alpha-[\beta+\delta])$ is calculable and is given by 
\bea
\alpha-[\beta+\delta]&=&\frac{1}{2}\sin^{-1}\left\lbrace\frac{I_{23}m_2^2s_{12}^4}{-\kappa m_3^2}\right\rbrace
\eea
along with an explicit relationship between the three invariants as
\bea
\frac{I_{23}}{c_{23}^2I_{12}+s_{23}^2I_{13}}\simeq -\frac{m_3^2}{m_2^2\sin^4 \theta_{12}} 
.\eea
Therefore, essentially we get two independent invariants corresponding to the Majorana phase difference and the Dirac CP phase.\\

\noindent
\textbf{Case III:} $m_2,m_3\neq0$, $m_1$, $\theta_{13}= 0$: {\B$One$ $independent$ $invariant$.}\\

\noindent
In such a condition the three invariants are coming out in a correlated manner as
\bea
I_{12}&=&\kappa \sin 2(\alpha-[\beta+\delta])\nonumber \\
&=&-\sin 2(\alpha-[\beta+\delta])c_{12}^2c_{23}^2s_{23}^2c_{13}^6s_{12}^4m_2^3m_3\nonumber \\
&=&I_{13}\nonumber \\
I_{23}&=&\sin 2(\alpha-[\beta+\delta])c_{12}^2c_{23}^2s_{23}^2c_{13}^6m_2m_3^3\nonumber \\
&=&\left(-\frac{m_3^2}{m_2^2}s_{12}^{-4}\right)I_{12}\nonumber \\
&=&\left(-\frac{m_3^2}{m_2^2}s_{12}^{-4}\right)I_{13}
\eea
and in this case only independent invariant $I_{12}$ is connected to the Majorana phase difference $(\alpha-[\beta+\delta])$.
\subsection{Quasi-degenerate case}\label{qsi}
Although in the present work we are not discussing the quasi-degenerate  case which is relevant in the cosmological context \cite{Beutler:2014yhv}, however, one can calculate the Majorana phases in a model independent way by directly solving the invariants as mentioned at the beginning of Sec. \ref{s4}. To be precise, using Eqn. (\ref{I1}), (\ref{I2}) and Eqn. (\ref{I3}) one can extract all the CP violating phases without any hierarchical assumption. However, the calculation is tedious and will be studied elsewhere. Another alternative way is to follow the calculations presented in Ref. \cite{Adhikary:2013bma}. However, in that case the phase convention is different. Utilizing the phase convention presented in this work one can calculate all the phases in the second method also.
\section{Numerical estimation}\label{s5}
\subsection{Parametrization, diagonalization and the ranges of the Majorana phases}
\begin{figure}[h!]
\begin{center}
\includegraphics[scale=.6]{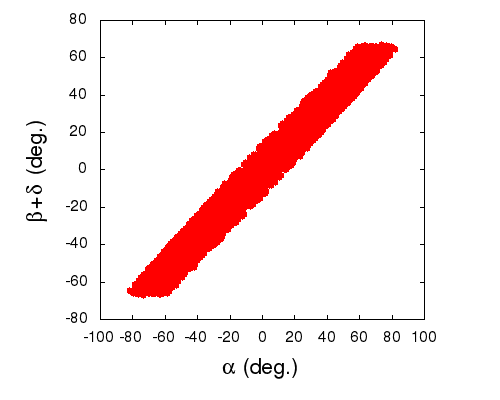}   \includegraphics[scale=.6]{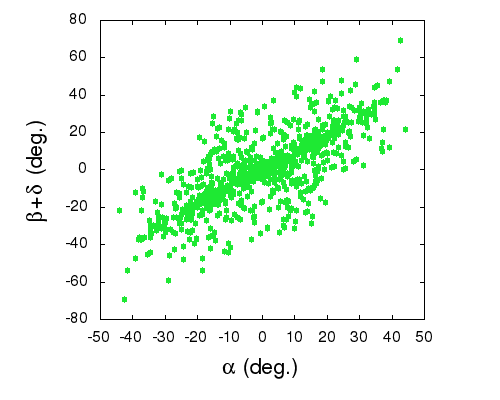} 
\end{center}
\caption{Plots of the Majorana phases ($\alpha$ vs $\beta+\delta$) for normal (left) and inverted (right) hierarchies .}\label{fig}
\end{figure}

A general solution for a three generation complex symmetric Majorana mass matrix is given in Ref.\cite{Adhikary:2013bma}. In order to estimate the Majorana phases obtained in the present work we utilize the expressions of the three eigenvalues and the three mixing angles. We also use the global fit data of neutrino oscillation experiments shown in Table \ref{tl} and the upper limits on the sum of the neutrino masses ($\Sigma_i m_i(=m_1+m_2+m_3)<0.23$ eV)\cite{Ade:2015xua} and the $\beta\beta_{0\nu}$ parameter ($|m_{11}|<0.35$ eV)\cite{Auger:2012ar} to obtain model independent ranges of the Majorana phases.\\
 \begin{table}[!h]
 \caption{Input experimental values\cite{Tortola:2012te}}\label{tl}
 \begin{center}
 \begin{tabular}{|c|c|}
\hline 
Quantity & 3$\sigma$ ranges \\ 
\hline 
$|\Delta m_{31}^2|$ (N)& 2.31$< \Delta m_{31}^2(10^3 eV^{-2})<2.74$ \\ 
\hline 
$|\Delta m_{31}^2|$ (I)& 2.21$< \Delta m_{31}^2(10^3 eV^{-2})<2.64$ \\ 
\hline 
 $\Delta m_{21}^2$& 7.21$< \Delta m_{21}^2(10^5 eV^{-2})<8.20$ \\ 
\hline 
$\theta_{12}$ & $31.3^o<\theta_{12}<37.46^o$ \\ 
\hline 
$\theta_{23}$ &  $36.86^o < \theta_{23}<55.55^o$  \\  
\hline 
$\theta_{13}$ &  $7.49^o < \theta_{13}< 10.46^o$  \\ 
\hline 
\end{tabular} 
\end{center} 
\end{table} \\
We consider a most general $3\times3$ complex symmetric neutrino mass matrix $m_\nu$ as 
\bea
m_\nu &=& \begin{pmatrix}
P&Q&R\\Q&S&T\\R&T&V
\end{pmatrix}
\eea
with all parameter complex and can be parametrized as 
\bea
m_\nu &=& m_0 e^{i\alpha_m}\begin{pmatrix}
1&xe^{i\alpha_x}&ye^{i\alpha_y}\\
xe^{i\alpha_x}&ze^{i\alpha_z}&we^{i\alpha_w}\\
ye^{i\alpha_y}&we^{i\alpha_w}&ve^{i\alpha_v}
\end{pmatrix}\label{mnug}
\eea
with the definition of the parameters 
\bea
P=m_0e^{i\alpha_m}, Q/P=xe^{i\alpha_x},R/P=ye^{i\alpha_y},S/P=ze^{i\alpha_z},T/P=we^{i\alpha_w},V/P=ve^{i\alpha_v}.
\eea
We can now give a phase rotation to the matrix of Eqn.(\ref{mnug}) by a diagonal phase matrix $K=diag(e^{i\phi_1},e^{i\phi_2},e^{}i\phi_3)$ as 
\bea
m_\nu^{\prime} &=& K^T m_\nu K
\eea
and consequently the rotated matrix comes out with 9 parameters as 
\bea
m_\nu^{\prime} &=& m_0\begin{pmatrix}
1&x&y\\
x&ze^{i\Omega_1}&we^{\Omega_2}\\
y&we^{i\Omega_2}&ve^{i\Omega_3}
\end{pmatrix}
\eea
where x,y,z,w are the real parameters and $\Omega_1$, $\Omega_2$, $\Omega_3$, $\phi_1$, $\phi_2$, $\phi_3$ are defined as 
\bea
\Omega_1=\alpha_z-2\alpha_x,\Omega_2=\alpha_w-\alpha_x-\alpha_y,\Omega_3=\alpha_v-2\alpha_y
\eea
and
\bea
\phi_1=-\frac{\alpha_m}{2},\phi_2=-(\alpha_x-\frac{\alpha_m}{2}),\phi_3=-(\alpha_y-\frac{\alpha_m}{2}).
\eea
Now using Eqn.(\ref{pi2}) we can explicitly calculate the rephasing invariants in terms of the elements of $m_\nu^{\prime}$. It is to be noted, that in the general case the number of parameters are 9 and we have only 7 experimental inputs. However, among the 9 parameters there are three angle parameters ($\Omega_1$, $\Omega_2$ and $\Omega_3$). We set the values of these angle parameters in an arbitrary manner within the range $0-2\pi$ and vary the other parameters in a wide range to estimate the  overall ranges of the Majorana phases which are depicted in figure \ref{fig}. We first constrain the rephasing invariants which in turn generate the correlated plot of the Majorana phases. The correlation  between the phases are the  consequences of Eqn.(\ref{c2}) and Eqn.(\ref{c1}) respectively.
 
\noindent
Upon numerical estimation,  the model independent ranges for $\alpha$ and $\beta+\delta$ come out as $-90^o<\alpha<90^o$ and $-71^o<\beta+\delta<71^o$ for normal hierarchy ($m_1\neq0$, $\theta_{13}\neq0$) and $-45^o<\alpha<45^o$, $-70^o<\beta+\delta<70^0$ for inverted hierarchy ($m_3\neq0$, $\theta_{13}\neq0$) and are shown explicitly in figure \ref{fig}. For $m_3=0$ case, the range of $\alpha$ is obtained as $-45^o<\alpha<45^o$ and for case $m_1=0$,  the phase difference is constrained as $-82^o<\alpha-[\beta+\delta]<82^o$. We also present the parameter ranges in Table \ref{tpr}.
\begin{table}
\begin{center}
\caption{Parameter ranges for a phenomenologically viable $m_\nu$.}\label{tpr}
\begin{tabular}{|p{1.3cm}|p{2.4cm}|p{2.4cm}|p{2.4cm}|p{2.4cm}|p{2.1cm}|p{2.1cm}|}
\cline{2-7}
\multicolumn{1}{c|}{{\B Hierarchies $\downarrow$}}&$m_0$&x&y&z&w&v\\
\hline
$Normal$ $m_1\neq0$&$2.4\times10^{-4}-1.7\times 10^{-3}$&$0.15<x<3.7$&$0.15<y<4.6$&$0.14<z<9.5$&$0.1<w<8.6$&$0.13<v<8.4$\\
\hline
$Normal$ $m_1=0$ &$2\times10^{-4}-1.2\times 10^{-3}$&$0.1<x<3.2$&$0.14<y<4.7$&$0.09<z<7.5$&$0.09<w<8$&$0.11<v<8.1$\\
\hline
$Inverted$ $m_3\neq0$ &$1.2\times10^{-4}-1.8\times 10^{-3}$ &$0.5<x<3.5$&$0.5<y<3.47$&$0.1<z<2.6$&$0<w<1.8$&$0<v<2.4$\\
\hline
$Inverted$ $m_3=0$ &$1.1\times10^{-4}-1.4\times 10^{-3}$ &$0.1<x<3$&$0.2<y<3.4$&$0<z<2.4$&$0<w<1.7$&$0<v<2.4$\\
\hline
\end{tabular}
\end{center}
\end{table}\\
\subsection{Connection to the physical observables and future of the Majorana phases}
\begin{figure}[h!]
\begin{center}
\includegraphics[scale=.6]{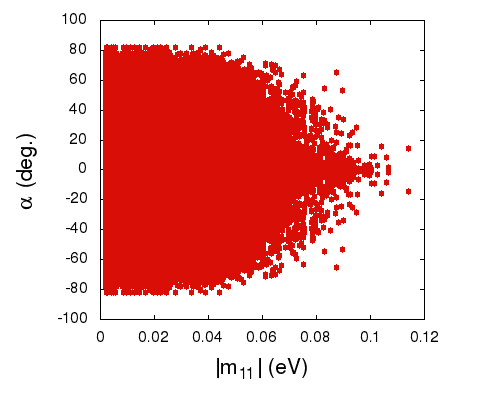}   \includegraphics[scale=.6]{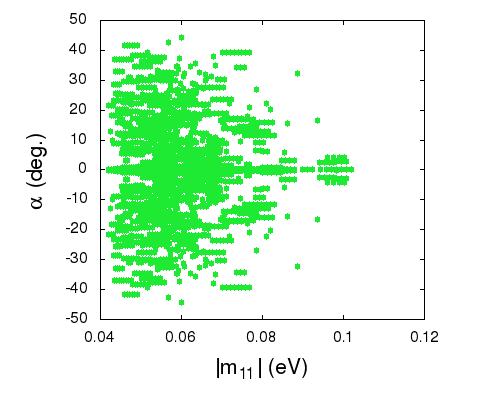}\\
\includegraphics[scale=.6]{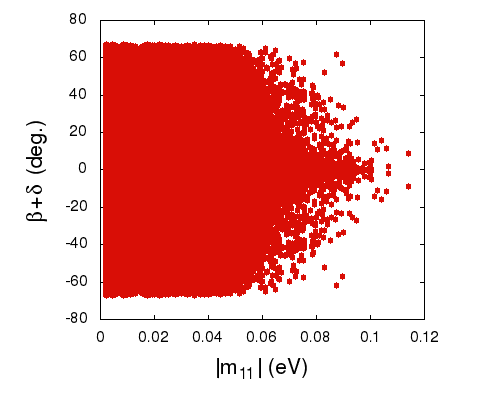}   \includegraphics[scale=.6]{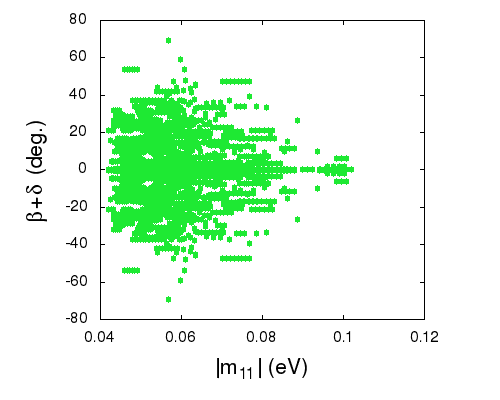}  
\end{center}
\caption{Plots of the Majorana phases ($\alpha$, $\beta+\delta$) vs $|m_{11}|$ for normal (left) and inverted (right) hierarchies  for best fit values of $\Delta m_{21}^2$.}\label{obs}
\end{figure}

As previously mentioned, unlike the Dirac CP phase $\delta$, the Majorana phases do not appear in the neutrino$\rightarrow$neutrino oscillation. Therefore, a natural question arises how and where these phases can be measured. As a direct detection, in Ref. \cite{Xing:2013ty} Xing  suggested a thought experiment (neutrino$\rightarrow$ antineutrino oscillation) in which he pointed out these phases may  appear in the probability expression of the flavour oscillation and thus also in the expression of the CP asymmetry parameter $\mathcal{A}_{\alpha\beta}$ which is the measure of CP violation. However, this kind of experiment is purely academic at this moment and practically difficult to design as the oscillation probability is highly suppressed by the factor $m_i^2/E^2$, where $m_i$ is the mass of the light neutrino and $E$ is the beam energy. Now considering $E$ $\sim$ MeV and the masses of the neutrinos to be less than 1 eV, one can calculate  $m_i^2/E^2$ to be $\mathcal{O}(10^{-12})$. To improve $m_i/E$,  a novel suggestion\cite{Xing:2013woa,Xing:2013ty} is to lower the value of $E$, however, in that case the estimated  size of the base line length  and the detector are beyond the reach of the present experimental facilities. But as an optimistic point of view we expect these kind of experiments will be designed in future and thus the prediction of the Majorana phases will be tested. Beside neutrino $\rightarrow$ antineutrino oscillation there are several LNV processes like $\beta\beta_{0\nu}$ decay, $\Delta^{++}\rightarrow l_{\alpha}^+ l_{\beta}^+$ (in Type II seesaw model)\cite{Xing:2013woa} etc., which play a crucial role for the indirect measurement of the Majorana phases. Now coming into our work, we present a table in the appendix which shows the ranges of the obtained Majorana phases for some typical values of $|m_{11}|$ and for convenience, in figure \ref{obs} we present variation of the Majorana phases with $|m_{11}|$ for the best fit value of $\Delta m_{21}^2$ and taking all the other constraints in their 3$\sigma$ ranges for both the hierarchies. We would like to mention that even if we take the 3$\sigma$ range of $\Delta m_{21}^2$, over all ranges of the Majorana phases do not differ much, however, unlike the plots of figure \ref{obs}, the plots in that case become more wider  for the higher values of $|m_{11}|$ ($>0.08$ eV). Although, the present experimental upper bound on $|m_{11}|$ is 0.35 eV, NEXT will be able to bring down the  value to 0.1 eV and thus the approximate ranges of the Majorana phases can be predicted. 

 Thus far we have estimated the Majorana phases in a general context. Latter, we apply the expressions obtained for  $\alpha$ and $\beta+\delta$ for few testable flavour models (models with lesser number of parameters) as an application of the general result although our analysis is true for any hierarchical model of neutrino masses.
 
\section{Some testable flavour models}\label{s6} 
The reason we discuss this section is to make certain whether the results obtain in the general case are consistent with the other models or not. Moreover the models with certain flavour symmetries are highly predictive in nature. Therefore, precise measurement of the CP violating phases may act as the important tools to verify the testability of the flavour models\cite{Shimizu:2015tta}.  In inverted hierarchy section we present a model with scaling ansatz and texture zeros within the framework of inverse seesaw through which all the sub cases presented in Sec. \ref{invh} can be realized while in the normal hierarchy section we present a model with cyclic symmetry within the framework of Type I seesaw. Obviously the choices are for illustration. One can also consider inverse or linear seesaw for normal hierarchy\cite{Adhikary:2013mfa,Chakraborty:2014hfa} and Type I seesaw for inverted hierarchy\cite{Adhikary:2012kb}. In principle one can use the technique in any hierarchical flavour models. The above numerical results are obtained for the general $m_\nu$ where all the 9 independent parameters are present. However, as previously said, one can reduce the number of parameters by invoking some symmetry or ansatz in the Lagrangian which is more predictive in nature and thus testable in the experiments. In this section we provide applications of the general results in few typical cases for both the hierarchies, normal and inverted.
\subsection{Normal hierarchy}
In this case we explore a model that corresponds to \textbf{Case I} of the normal hierarchical scenario mentioned in section (\ref{nrmlh}). The model is based on cyclic symmetry with Type I seesaw mechanism  to accommodate the neutrino oscillation data. In the fundamental level the symmetry exists in the neutrino sector of the Lagrangian and due to the symmetry a degeneracy in masses occurs removal of which  therefore requires breaking of the symmetry. It is shown that a minimal breaking in the Majorana mass matrix is sufficient to explain the extant data. In this model the low energy broken symmetric mass matrix $m_\nu(=-m_DM_R^{-1}m_D^T)$ originated from Type I seesaw mechanism is given by
\bea
m_\nu = m_0 \begin{pmatrix}
p^2e^{2i\alpha}+\frac{q^2e^{2i\beta}}{1+\frac{\epsilon_1}{m}}+\frac{1}{1+\frac{\epsilon_2}{m}}&pe^{i\alpha}+\frac{pqe^{i(\alpha+\beta)}}{1+\frac{\epsilon_1}{m}}+\frac{qe^{i\beta}}{1+\frac{\epsilon_2}{m}}& \frac{pe^{i\alpha}}{1+\frac{\epsilon_2}{m}}+pqe^{i(\alpha+\beta)}+\frac{qe^{i\beta}}{1+\frac{\epsilon_1}{m}}\\pe^{i\alpha}+\frac{pqe^{i(\alpha+\beta)}}{1+\frac{\epsilon_1}{m}}+\frac{qe^{i\beta}}{1+\frac{\epsilon_2}{m}}&1+\frac{p^2e^{2i\alpha}}{1+\frac{\epsilon_1}{m}}+\frac{q^2e^{2i\beta}}{1+\frac{\epsilon_2}{m}}&\frac{pe^{i\alpha}}{1+\frac{\epsilon_1}{m}}+\frac{pqe^{i(\alpha+\beta)}}{1+\frac{\epsilon_2}{m}}+qe^{i\beta}\\ \frac{pe^{i\alpha}}{1+\frac{\epsilon_2}{m}}+pqe^{i(\alpha+\beta)}+\frac{qe^{i\beta}}{1+\frac{\epsilon_1}{m}}&\frac{pe^{i\alpha}}{1+\frac{\epsilon_1}{m}}+\frac{pqe^{i(\alpha+\beta)}}{1+\frac{\epsilon_2}{m}}+qe^{i\beta}&\frac{p^2e^{2i\alpha}}{1+\frac{\epsilon_2}{m}}+q^2e^{2i\beta}+\frac{1}{1+\frac{\epsilon_1}{m}}
\end{pmatrix}
\eea
where
\bea
m_D =\begin{pmatrix}
y_1&y_2&y_3\\y_3&y_1&y_2\\y_2&y_3&y_1
\end{pmatrix},
\eea
 \bea
M_R=diag(m+ \epsilon_1, m+\epsilon_2, m)
\eea
with  $\epsilon_1$ and $\epsilon_2$ as the breaking parameters and 
\bea
m_0=-\frac{y_3^2}{m},p e^{i\alpha}=\frac{y_1}{y_3},q e^{i\beta}=\frac{y_2}{y_3}.
\eea
\begin{figure}[h!]
\begin{center}
\includegraphics[scale=.55]{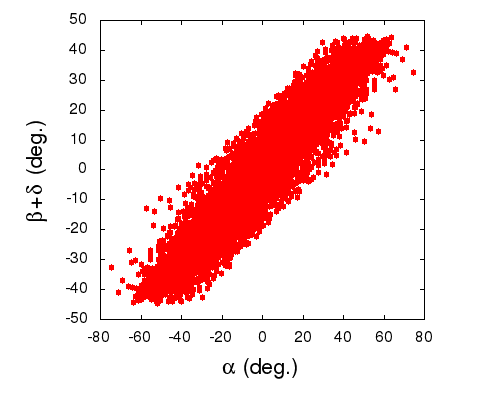} 
\caption{Correlation of $\alpha$ vs $\beta+\delta$.}\label{fig1}
\end{center}
\end{figure}\\
 For numerical analysis we choose the mass scale of $M_R$ to be of the order of $10^{15}$ GeV and $m_D$ to be at electroweak scale. Further redefining the breaking parameters as $\epsilon_1^\prime=\frac{\epsilon_1}{m}$ and $\epsilon_2^\prime=\frac{\epsilon_2}{m}$ we allow them to vary as $-0.1<\epsilon^\prime_1,\epsilon^\prime_2<0.1$ to keep the breaking effect small. We then constrain the parameter spaces taking into account the 3$\sigma$ ranges of  neutrino oscillation global fit data and  explicitly evaluate both the Majorana phases. From figure \ref{fig1}  the ranges read as $-77.2^o<\alpha<76.7^o$ and $-45.3^o<\beta+\delta<45.5^o$. Note that the ranges of both the phases are embedded within the values obtained for the general case. Similar to the general case, in figure \ref{obsn} we also present the variation of the Majorana phases with the $\beta\beta_{0\nu}$ parameter.  One can see the upper limit of $|m_{11}|$ is $\sim$ 0.07 eV which is well within the reach of the future planned experiments.\\
\begin{figure}[h!]
\begin{center}
\includegraphics[scale=.6]{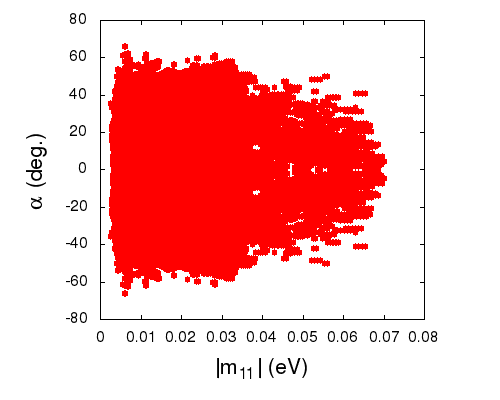}   \includegraphics[scale=.6]{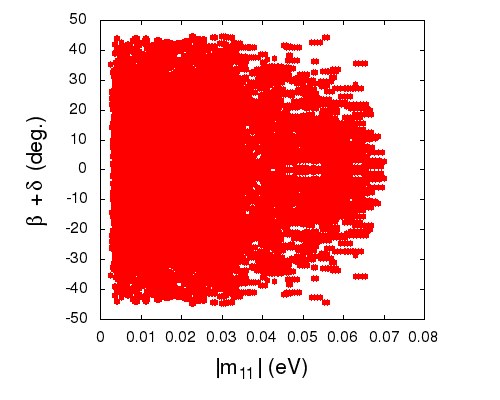}\\
\caption{Variation of $\alpha$ and $\beta+\delta$ with $|m_{11}|$ for cyclic symmetric case (normal hierarchy).}\label{obsn}
\end{center}
\end{figure}
As the model consists of lesser number of parameters, we also expect a significant correlation between the phase invariants and are depicted in figure \ref{incor1}.
\begin{figure}[h!]
\begin{center}
\includegraphics[scale=.6]{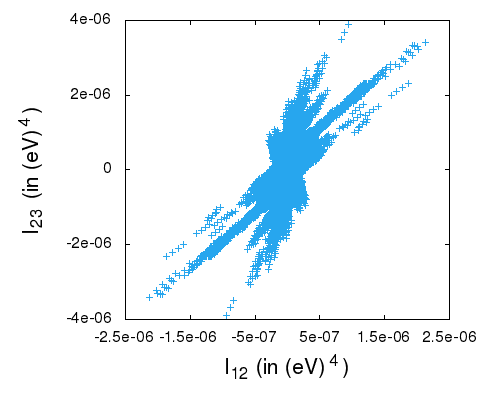}   \includegraphics[scale=.6]{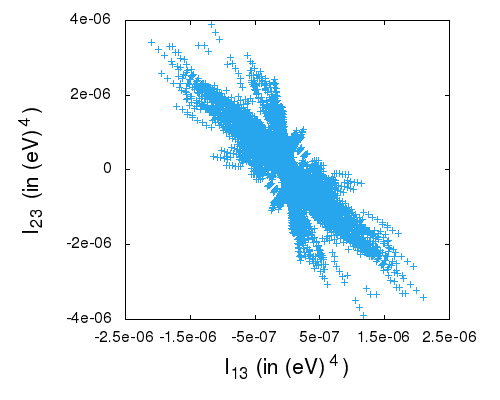}\\
\caption{Correlated plots of the rephasing invariants, $I_{12}$ vs $I_{23}$ (left) and $I_{13}$ vs $I_{23}$ (right).}\label{incor1}
\end{center}
\end{figure}
\subsection{Inverted hierarchy}
\begin{figure}[h!]
\begin{center}
\includegraphics[scale=0.6]{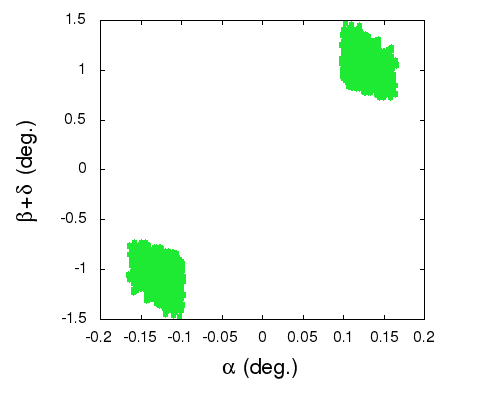}\\
\includegraphics[scale=.6]{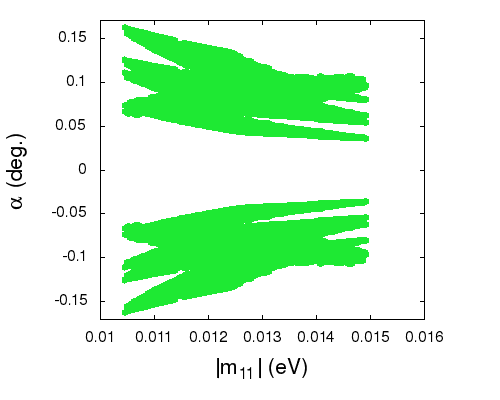}   \includegraphics[scale=.6]{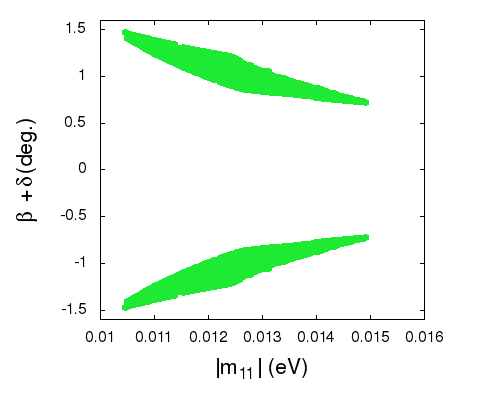}
\caption{Correlation plot of $\alpha$ vs $\beta+\delta$ (upper panel) and variation of $\alpha$ and $\beta+\delta$ with $|m_{11}|$ for inverted hierarchy : scaling ansatz case (lower panel).}\label{npb}
\end{center}
 \end{figure}

\begin{figure}[h!]
\begin{center}
\includegraphics[scale=.55]{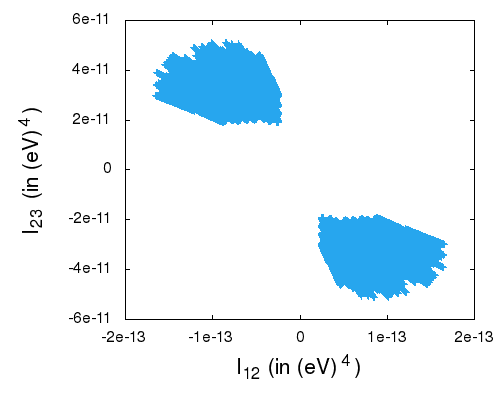}   \includegraphics[scale=.55]{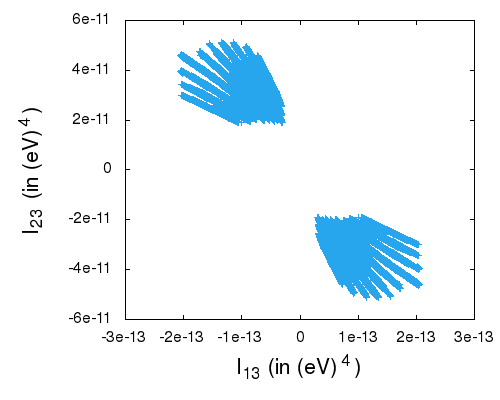}
\caption{Correlated plots of the rephasing invariants, $I_{12}$ vs $I_{23}$ (left) and $I_{13}$ vs $I_{23}$ (right).}\label{incor2}
\end{center}
\end{figure}

In this case , we explore a model based on scaling ansatz with inverse seesaw mechanism\cite{ Mohapatra:1986bd,Bernabeu:1987gr,Mohapatra:1986aw,Schechter:1981cv,Dev:2009aw,Awasthi:2013ff,Abada:2014vea,Palcu:2014aza,Schechter:1980gr,Fraser:2014yha,Hettmansperger:2011bt,Adhikary:2013mfa,Law:2013gma} . In this mechanism $m_\nu$ is given by 
\bea
m_\nu &=& m_D M_{RS}^{-1}\mu ( m_D M_{RS}^{-1})^T \nonumber \\ 
\eea
where $m_D$ is the usual Dirac type matrix and the other two matrices $\mu$ (Majorana type) and $M_{RS}$ (Dirac type) arise due to the interaction between the additional singlet fermion and right handed neutrino considered in this type of seesaw mechanism. To further reduce the number of parameters texture zeros\cite{Blum:2007qm,Obara:2007nb,Damanik:2007yg,Grimus:2004cj,Dev:2013esa,Adhikary:2012kb,Frampton:2002yf,
Whisnant:2014yca,Ludl:2014axa,Ludl:2015lta,Grimus:2013qea,Lavoura:2015wwa,
Ferreira:2014vna,Xing:2015sva,Xing:2002ta,Liao:2013saa,Fritzsch:2011qv,Merle:2006du,
Wang:2013woa,Wang:2014dka,Lavoura:2004tu,Kageyama:2002zw,
Wang:2014vua,4zero1,Choubey:2008tb,Chakraborty:2014hfa,4zero2,4zero3,4zero4,Ghosal:2015lwa}  are assumed in the constituent $m_D$ and $\mu$ matrices. Scaling ansatz invariance dictates $m_3=0$ and $\theta_{13}=0$ and this case corresponds to \textbf{Case III} of Sec. \ref{invh}. Thus to generate non zero $\theta_{13}$ breaking of the ansatz is necessary. Incorporating breaking in $m_D$ through a small parameter $\epsilon$, there are two different phenomenologically survived textures which are given by 
\bea
 m_\nu^{1}&=& m_0\begin{pmatrix}
1& k_1p & p \\
k_1p  & k_1^2(q^2e^{i\theta}+p^2)& k_1(q^2e^{i\theta}+p^2)\\p & k_1(q^2e^{i\theta}+p^2)&(q^2e^{i\theta}+p^2)
\end{pmatrix}+m_0 \epsilon \begin{pmatrix}
0&0&0\\0&2k_1^2q^2e^{i\theta}&k_1q^2e^{i\theta}\\0&k_1q^2e^{i\theta}&0\\
\end{pmatrix}\label{mAB}  
\eea
and 
\bea
m_\nu^{2} = m_0 \begin{pmatrix}
 1&k_1(p+qe^{i\theta})&p+qe^{i\theta}\\
 k_1(p+qe^{i\theta})&k_1^2(2pqe^{i\theta}+p^2)&k_1(2pqe^{i\theta}+p^2)\\
 p+qe^{i\theta}&k_1(2pqe^{i\theta}+p^2)&(2pqe^{i\theta}+p^2)\\
 \end{pmatrix} \nonumber \\+m_0 \epsilon \begin{pmatrix}
 0&k_1qe^{i\theta}&0\\k_1qe^{i\theta}&2k_1^2pqe^{i\theta}&k_1pqe^{i\theta}\\
 0&k_1pqe^{i\theta}&0\\
 \end{pmatrix}\label{mc} 
 \eea
 where all the parameters are complex \cite{Ghosal:2015lwa}. In both the cases $\theta_{13}\neq0$ however, $m_3=0$ due to singular nature of $\mu$ matrix and this case corresponds to \textbf{Case II} of Sec. \ref{invh}.\\
 \noindent
 We further consider the most general version of the above case through the breaking of the ansatz in both $m_D$ and $\mu$ matrices through two small parameters $\epsilon$ and $\epsilon^\prime$ respectively and the neutrino mass matrix $m_{\nu}^3$ comes out as 
 \bea
 m_{\nu }^{3}= m_0\begin{pmatrix}
1& k_1p & p \\
k_1p  & k_1^2(q^2e^{i\theta}+p^2)& k_1(q^2e^{i\theta}+p^2)\\p & k_1(q^2e^{i\theta}+p^2)&(q^2e^{i\theta}+p^2)
\end{pmatrix}+m_0 \epsilon \begin{pmatrix}
0&0&0\\0&2k_1^2q^2e^{i\theta}&k_1q^2e^{i\theta}\\0&k_1q^2e^{i\theta}&0\\
\end{pmatrix} \nonumber \\ +m_0 \epsilon^\prime \begin{pmatrix}
0&k_1p&p\\k_1p&0&0\\p&0&0
\end{pmatrix}\label{mnu3}
\eea 
and in this situation both $\theta_{13}$ and $m_3$ are nonzero corresponding to \textbf{Case I} of Sec. \ref{invh}. Thus the whole inverted hierarchical sector is generated through the choice of the above model.\\
\noindent
Now, with the explicit construction of rephasing invariants we calculate the Majorana phases in each case. Interestingly, for all the cases, the value of $J_{CP}$ comes out very small due to smallness of the Dirac CP phase $\delta$, or more precisely, due to almost real nature of the mass matrices. Therefore, such typical nature of the mass matrices also constrains the Majorana phases approximately as $-1.2^o<\alpha<0.8^o$ for the first two matrices ($m_\nu^1$ and $m_\nu^2$) and $-0.17^o<\alpha<0.17^o$,$-1.5^o<\beta+\delta<1.5^o$ for the matrix $m_\nu^3$ along with an approximate range of $\beta\beta_{0\nu}$ parameter $|m_{11}|$ as 0.01 eV $<|m_{11}|<$ 0.0148 eV and 0.01 eV $<|m_{11}|<$ 0.0152 eV respectively. For illustration, in figure \ref{npb} we plot $\alpha$ and $\beta+\delta$ with $|m_{11}|$ for $m^3_\nu$. For other two matrices ($m^1_\nu$ and $m^2_\nu$) the variations of $\alpha$ with $|m_{11}|$ are almost same as that of the extreme left plot of the lower panel of figure \ref{npb}. The model is highly predictive and hence, if significant CP violation is observed, the model will be ruled out.
 We plot the correlation between the invariants in figure \ref{incor2}.\\
\noindent 
 Finally, we summarize our results in Table \ref{t2} that shows the ranges of the Majorana phases for all the cases.
 \begin{table}[!h]
\caption{Summary of the numerical results.}\label{t2}
\begin{center}
\begin{tabular}{|m{3.0cm}|m{2.0cm}|m{2.0cm}|m{2.5cm}|m{2.0cm}|m{2.0cm}|}
\cline{2-6}
\multicolumn{1}{c|}{}  & \multicolumn{2}{c|}{\textbf{General case}} & \multicolumn{1}{c|}{\textbf{Cyclic symmetry}} & \multicolumn{2}{c|}{\textbf{Scaling ansatz }}\\ 
\cline{2-6}
\multicolumn{1}{c|}{\G{Hierarchies $\rightarrow$}} & $Normal$ & $Inverted$ & $Normal$ & $Inverted$ & $Inverted$\\
\cline{2-6}
\multicolumn{1}{c|}{\B{Particular cases $\rightarrow$}} & $m_1\neq 0$ $\theta_{13}\neq 0$ & $m_3\neq 0$ $\theta_{13}\neq 0$ & $m_1\neq 0$ $\theta_{13}\neq 0$ &  $m_3\neq 0$ $\theta_{13}\neq 0$ &  $m_3= 0$ $\theta_{13}\neq 0$\\
\hline
 $\alpha$ ( deg.) & $-90-90$ & $-45-45$ &  $-77.2-76.7$ & $-0.17-0.17$ & $-1.2-0.8$\\
\hline
$\beta+\delta$ ( deg.)&$-71-71$&$-70-70$&$-45.3-45.5$&$-1.5-1.5$&$absent$\\
\hline
\end{tabular}
\end{center}
\end{table} 
 \section{Summary}\label{s7}
 In the present work we calculate the Majorana phases of a general complex symmetric $3\times3$ neutrino mass matrix utilizing the three rephasing invariant quantities $I_{12}$, $I_{13}$ and $I_{23}$ proposed by Sarkar and Singh  for both the hierarchical structures of neutrinos using Mohapatra-Rodejohann's phase convention. Motivation behind the usage of the invariants to calculate the Majorana phases is that such methodology enables us to evaluate the existing Majorana phase even if one of the eigenvalue ($m_3$) is zero in a model independent way. However, if $m_1=0$, this methodology will only enable us to calculate the difference of the Majorana phases therefore it is needed to change the phase convention in that case. After the presentation of the generalized prescription, we further  present  the maximal allowed ranges of the Majorana phases in general context for both the viable hierarchical structures of neutrino masses and address our methodology to be true for any model except the case of quasi degeneracy. We  then talk about the connection of the Majorana phases with physical observables like $\beta\beta_{0\nu}$ parameter $|m_{11}|$ and the branching ratios of charged Higgs ($\Delta^{++}$) decay where the phases show up. As a direct measurement of the Majorana phases we give the example of neutrino$\rightarrow$ antineutrino oscillation which is a thought experiment right now, however, well studied in literature. Although the presented methodology can be used in any hierarchical models in neutrino physics, after discussing the general case  we further exemplify our methodology  in few typical testable models (models with lesser number of parameters) leading to normal and inverted hierarchy and with their significant predictions on $|m_{11}|$ and the Majorana phases. For normal hierarchical case we give an example of a model based on cyclic symmetry with Type I seesaw mechanism. We estimate the Majorana phases for the broken symmetric case, since cyclic symmetry dictates a degeneracy in the mass eigenvalues. As an example of inverted hierarchy, we cite a  model with high predictability and comprised of scaling ansatz, texture zeros and inverse seesaw mechanism. It is seen that all the sub cases of inverted hierarchy  mentioned in Section \ref{invh} can be obtained  depending upon the scheme of incorporation of ansatz breaking mechanism  while a phenomenologically viable sub case ($m_1=0$) of the normal hierarchy  is yet to be established through the choice of a suitable model.
 \newpage
 \appendix
 \section{Appendix}
 \begin{table}[!h]
\caption{ Majorana phases for the general $m_\nu$ for some typical values of $|m_{11}|$.}\label{sym}
\begin{center}
\begin{tabular}{|p{1.5cm}|p{1cm}|p{1cm}|p{1cm}|p{1cm}|p{1cm}|p{1cm}|p{1cm}|p{1cm}|p{1cm}|p{1cm}|}
\hline
\multicolumn{11}{|c|}{$Normal$ $Hierarchy$}\\ 
\hline
\multicolumn{1}{c|}{$|m_{11}|$ (eV)$\rightarrow$} & $0.001$ & $0.005$ & $0.01$ & $0.05$ & $0.1$ &$0.12$& $0.15$ & $0.2$ & $0.25$&$0.30$ \\
\hline
$\alpha$ (deg.) & $-90-90$&$-90-90$ & $-87-87$ & $-82-82$ & $-80-80$ & $-77-77$ & $-76-76$ & $-75-75$ & $-75-75$ & $-74-74$ \\
\hline
$\beta+\delta $ (deg.) & $-71-71$ & $-71-71$ &$-71-71$ & $-70-70$ & $-70-70$ &  $-69-69$&  $-67-67$ & $-65-65$ &  $-65-65$ & $-64-64$\\
\hline
\hline
\hline
\multicolumn{11}{|c|}{$Inverted$ $Hierarchy$}\\ 
\hline
\multicolumn{1}{c|}{$|m_{11}|$ (eV)$\rightarrow$}  &$0.01$& $0.03$ & $0.05$ & $0.08$ & $0.10$ & $0.12$ & $0.15$ & $0.2$ & $0.25$&$0.30$ \\
\hline
$\alpha$ (deg.) & $-70-70$ & $-70-70$ &$-70-70$ & $-69-69$ & $-69-69$ &  $-69-69$&  $-65-65$ & $-65-65$ &  $-63-63$ & $-63-63$ \\
\hline
$\beta+\delta $ (deg.)& $-46-46$ & $-46-46$ &$-46-46$ & $-45-45$ & $-45-45$ &  $-44-44$&  $-43-43$ & $-42-42$ &  $-42-42$ & $-40-40$\\
\hline
\end{tabular}
\end{center}
\end{table} 
\noindent
\textbf{Acknowledgement}\\

 We thank Prof. Probir Roy for careful reading and useful comments.
 {}
\end{document}